\documentclass[aps,pre,superscriptaddress,floatfix,12pt]{revtex4-1}

\usepackage{amsmath,amsfonts,amssymb}
\usepackage{graphicx}
\usepackage{algorithm}
\usepackage{algorithmic}
\usepackage{epstopdf}
\usepackage[hyperfigures,breaklinks,colorlinks,citecolor=blue,linkcolor=black]{hyperref}
\usepackage{subfig}
\usepackage{mathbbol}
\def\s{\sigma}
\def\ul{\underline}
\def\p{\partial}

\newcommand{\lasubsim}[3]{\lambda_{#1 \rightarrow (#1 #2)} (#3)}
\newcommand{\musubsim}[3]{    \mu_{#1 \rightarrow (#1 #2)} (#3)}
\newcommand{\musub}[3]{    \mu_{#1 \rightarrow (#1 #2)} (X_{#1}(#3)|X_{#2}(#3))}
\newcommand{\musubX}[2]{    \mu_{#1 \rightarrow (#1 #2)}(X_{#1}|X_{#2})}

\begin{document}
\title{A Cavity Master Equation for the continuous time dynamics of  discrete spins models}

\author{E. Aurell} 
\affiliation{Department of Computational Biology, AlbaNova University Center, SE-106, 91 Stockholm, Sweden}
\affiliation{ Department of
Information and Computer Science, Aalto University, FIN-00076 Aalto, Finland}
\author{G. Del Ferraro} 
\affiliation{Department of Computational Biology, AlbaNova University Center, SE-106, 91 Stockholm, Sweden}
\author{E. Dom\'{\i}nguez}
\affiliation{Group of Complex Systems and Statistical Physics. Department of Theoretical Physics, Physics Faculty, University of Havana, Cuba}
\email{eduardo@fisica.uh.cu}
\author{R. Mulet}
\affiliation{Group of Complex Systems and Statistical Physics. Department of Theoretical Physics, Physics Faculty, University of Havana, Cuba}
\email{mulet@fisica.uh.cu}

\date{\today}

\begin{abstract}
\noindent
We present a new method to close the Master Equation representing the continuous time dynamics of Ising interacting spins. 
The method makes use of the the theory of Random Point Processes to derive a {\em master equation for local conditional probabilities}.
We analytically test our solution studying two known cases, the dynamics of the mean field ferromagnet and the dynamics of the one dimensional Ising system. We then present numerical results comparing our predictions with Monte Carlo simulations in three different models on random graphs with finite connectivity: the Ising ferromagnet, the Random Field Ising model, and the Viana-Bray spin-glass model.
 
\end{abstract}
\maketitle

\section{Introduction}

The comprehension of the properties of complex systems with many interacting particles is at the forefront of the research activity in many scientific communities, statistical physics \cite{kadanoff}, chemical kinetics\cite{ChemKin}, population biology\cite{PopBio}, neuroscience\cite{Neurosc}, etc. A very general approach to gain proper insight in these systems is to develop simple models that, although composed of many interacting particles, could be treated analytically and/or computationaly on reasonable time scales.

A prototypical example of this class of simple systems is the Ising ferromagnetic model whose equilibrium properties, despite in three dimensions still lack a proper analytical solution, are globally well understood\cite{Huang}. The addition of disorder to this model generates a more complicated scenario but, first the replica trick, together with the possibility of a replica symmetry breaking scheme \cite{MezardParisiVirasoro,parisi1980sequence,mezard1984replica}, and then the cavity approach introduced later \cite{mezard2003cavity,MezardParisi2001} - opened the doors to the analytical treatment of this and other models \cite{KS,COL,Weigt}. Provided the model is defined on fully connected or on general random graphs and leaving aside technical difficulties associated to each particular problem, there are proper tools to correctly understand the equilibrium properties of these families of disordered models \cite{mezard2009information}. Along this direction, only finite dimensional systems continue to be elusive albeit some progress have been obtained in the last few years \cite{Rizzo,JPhysA,Zhu11,Zhu12}.

The situation is completely different once the interest turns to the dynamical properties of complex systems. First because it is clear that approximations which work for short time scales are not necessarily valid at long time scales and viceversa. Second because regardless of the approximation, the selection of the local dynamical rules that define the processes turns out to be fundamental to characterize the evolution of the macroscopic quantities of the model \cite{gardiner1985handbook,van1992stochastic}. It is therefore not surprising that progresses in this direction have been slower. The introduction of disorder further complicates the issue, generating new families of models and behaviours that still lack a complete understanding \cite{cavagna}. 

As for the equilibrium, a first classification of complex systems in the dynamical scenario can be made between continuous and discrete state variables. The dynamical modelling of the former case is usually done by using a Langevin equation  \cite{van1992stochastic} as it is possible to write the differential of a spin variable in a rigorous form. For the same reason, in these models time is in most of the cases considered a continuous variable. The dynamics of the fully connected (FC) spherical $p$-spin model, for instance, has been studied in \cite{crisanti1993sphericalp} and solved by writing equations for the correlation and response function. The FC case $p=2$ has been considered in \cite{cugliandolo1995full} whereas random network architectures have been studied in \cite{semerjian2004stochastic} introducing a series of approximate equations. 

For discrete variables, the discontinuous nature of the spin values makes cumbersome, if not improper, to formulate the problem starting with Langevin-like equations. Thus in this case, instead of writing differential equations for the spin variables, one describes the stochasticity of the dynamics due to the contact with the heat bath by writing equations for the probability of the spin state. The literature has focused on two different possible choices for the dynamics: either time evolves in discrete steps or in continuous time. 

 Furthermore, due to the possible relevance of these models to understand the
dynamics in biological and neural networks, studies have focused also on cases with different network connectivity symmetries.  
The exact solution of a dilute fully asymmetric neural network model, for both parallel and asynchronous dynamics, dates back to \cite{derrida1987exactly}. Directed random graphs have been investigated by using different approaches. Path integral techniques for this case have been introduced in \cite{hatchett2004parallel} for parallel dynamics on graphs with finite connectivity. The tree-like structure of such graphs led then the extension of equilibrium techniques as the cavity method or belief-propagation message-passing \cite{mezard2009information} to the dynamical scenario. The pioneering contribution \cite{neri2009cavity}, followed by \cite{aurell2011message}, generalized the cavity technique to the parallel dynamics of the Ising model on a Bethe lattice but the approach was limited to the stationary solution of directed graphs. Random sequential updates rules have been considered in \cite{aurell2012dynamic} but also limited to stationary states. In \cite{kanoria2011majority} a similar technique has been applied to study models with majority dynamics update rules (\emph{i.e.} linear dynamics with thresholding) .  The first generalization to the out-of-equilibrium scenario of these techniques for models with reversible dynamics has been made in \cite{del2015dynamic}, where the authors introduced an approach to estimate single site and joint probability distributions for every time in the dynamics and for every connectivity symmetry. Followed then by \cite{barthel2015matrix} where further improvements of the dynamic reconstruction have been introduced at the price of a more involving formalism. More recently, novel variational approaches have shown to give accurate results for the transient dynamic of disordered spin systems by using a simple and systematic theory \cite{dominguez2016simple}.

The above historical overview shows that the field is active but also that progresses have been limited to discrete-time update rules. The continuous-time counterpart has indeed been investigated much less. In this case, a proper dynamic description of the spin state configuration, follows a Master Equation (ME) for the probability density of the states of $N$-spin interacting variables \cite{Glauber63, coolen2005theory}. Once the ME is defined one must select the rules of the dynamical evolution. Although this approach can be stated easily, the full solution of the master equation in the general case is a cumbersome task and exact solutions have been limited to simple models \cite{Glauber63,mayer2004general}. A way out to this issue has been proposed through the Dynamical Replica Analysis for fully connected \cite{laughton1996order} and diluted graphs \cite{mozeika2008dynamical} where the authors, instead that searching for an estimation of the out-of-equilibrium  probability of the spins state, derive dynamical equation for the probability of some macroscopic observables. Therefore if, on one side, this approach obviously reduces the dimensionality of the problem, on the other, it looses detailed information about the microscopic state of the system. 

In this contribution we focus on continuous-time dynamics for discrete-spin variables and derive a set of closed equations for marginal probabilities of the microscopic spin configuration.

The paper is organized as follows. In the next section we define the model in its more general form and present the main results of our contribution. Starting from the Master Equation of $N$-interacting discrete variables we derive, for models defined on random graphs, a master equation for the single spin probability that depends on local conditional probabilities.  In order to close this equation, we need an expression for the evolution of these latter probabilities. To achieve this goal, we introduce the formalism of Random Point Processes (RPP) and, considering a tree-like architecture of the graph,  we re-derive a known equation for conditional probabilities of spin histories, called dynamic message-passing (or dynamic cavity equation). Then, we obtain a new, now closed, master equation for the local conditional probabilities introduced above. We call this equation the {\em Cavity Master Equation} (CME); it represents the main formal result of our contribution. We then test the validity of this approach re-visiting two well known models for the Ising ferromagnet: the fully connected or mean-field case and one-dimensional  chain. Finally, we numerically test the performances of our approach for models defined on random graphs comparing our predictions with results from Monte Carlo simulations.

\section{The Model Dynamics}

We consider a system of $N$ interacting discrete spins variables $\ul\sigma = \{\sigma_1,\dots, \sigma_N\}$, with $\s_i = \pm 1$, in contact with a bath at a constant temperature that leads to spontaneous spin flips. In the more general scenario the transition rate $r_i(\ul\sigma)$ of having a spin flip for spin $i$ at time $t$  depends on all the spin variables in the system. Assuming that the probability of the configuration of the system at time $t$, $P (\ul\sigma,t)$, can be written in a Markovian form, a master equation describes the evolution of such probability in a continuous time formulation \cite{Glauber63, coolen2005theory}
\begin{equation}
\frac{d P (\ul\sigma) }{dt} = - \sum_{i=1}^N \Big[
r_i(\ul\sigma) P (\ul\sigma) - r_i(F_i(\ul\sigma)) P (F_i(\ul\sigma) ) \Big]\, ,
\label{eq:originalME}
\end{equation}
where we omitted the time dependence in $P(\ul\sigma,t)$ to shorten notation and $F_i$ represents the inversion operator on spin $i$, \emph{i.e.} $F_i (\ul\sigma) = \{\sigma_1, \dots, \sigma_{i-1}, -\sigma_{i},\sigma_{i+1}, \dots, \sigma_N\}$. Although (\ref{eq:originalME}) is a simple equation to state formally, in practice it implies the daunting task of tracking the evolution in time of $2^N$ discrete states. 

If the transition rate $r_i(\ul\sigma)$ does not depend on the whole system state but only on the configuration of spin $i$ and its neighbours $\partial i$, the master equation equation takes a more local form. In this case, the evolution in time of the probability of the spin configuration $\sigma_i$ is simply obtained tracing (\ref{eq:originalME}) over all the spin states except $\sigma_i$. The resulting equation reads
\begin{equation}
\frac{d P (\sigma_i) }{dt} = - \sum_{\sigma_{\p i}}  \Big[
r_i(\sigma_i, \sigma_{\p i}) P (\sigma_i,\sigma_{\p i} ) -  r_i(-\sigma_i, \sigma_{\p i}) P (-\sigma_i,\sigma_{\p i} ) \Big]
\label{eq:localME}
\end{equation}
\noindent where $\sigma_{\p i}$ stands for the configuration of all the spins in the neighbourhood of $i$. As for $P(\ul\sigma,t)$ in \eqref{eq:originalME}, in the equation above and hereafter we omit for brevity the time dependence of every probability distribution, all them should be considered being at time $t$.
 
At variance with \eqref{eq:originalME}, equation \eqref{eq:localME} although apparently simpler is not closed. On the left hand side we have the probability $P(\sigma_i)$ that spin $i$ is in a particular state whereas, on the right hand side,  $P(\sigma_i,\sigma_{\p i} )$ stands for the probability of a certain configuration for spin $i$ and its neighbours. To consistently describe the evolution of the single site probability \eqref{eq:localME} in time, we then have to search for a closure of this equation which is physically meaningful. From hereon, different approximations could be made on the joint probability  $P(\sigma_i,\sigma_{\p i} )$ appearing on the right hand side of \eqref{eq:localME}.  A reasonable one, that we will use through this work, is to assume that
\begin{equation}
P(\sigma_i,\sigma_{\p i} ) = \prod_{k \in \p i } P(\sigma_k | \sigma_i) P(\sigma_i)
\label{eq:factorP}
\end{equation}
\noindent which has the desirable property of being exact at equilibrium for trees and random graphs where loops are large compared to the system size. 
Assuming a tree-like topology and the factorization in \eqref{eq:factorP}, the master equation (\ref{eq:localME}) can then be written as
\begin{equation}
\frac{d P (\sigma_i) }{dt} = - \sum_{\sigma_{\p i}} \Big[
r_i(\sigma_i, \sigma_{\p i}) \big[ \prod_{k \in \p i }
P(\sigma_k| \sigma_i) \big] P(\sigma_i)
-  r_i(-\sigma_i, \sigma_{\p i}) 
\big[ \prod_{k \in \p i }
P(\sigma_k| -\sigma_i) \big]P(-\sigma_i) \Big]
\label{eq:localMEfact}
\end{equation}
The above equation is also not closed, as we do not know how $P(\sigma_k| \sigma_i)$ changes with time.
The knowledge of a closed equation for this latter probability would allow to describe the evolution of the probability distribution of the variable $\sigma_i$ and of the conditional probability $P(\sigma_k | \sigma_i)$ at each time in the dynamics. 

Next sections aims to derive a new  master equation for a conditional probability distribution 
 $p(\s_i |\s_k)$ that approximates $P(\s_i|\s_k)$ and under certain conditions is equal to it. As we will see, there are subtleties regarding
 the definition of conditional probabilities for interacting systems.  Our derivation is guided by very general probabilistic principles and as a result we obtain a closure scheme for $p(\s_i |\s_k)$. We here present the final result 
\begin{align}
 \nonumber
 \frac{d p(\s_i|\s_j)}{d t} = -  \sum_{\sigma_{\partial i\setminus j}} \Bigg[& r_{i}[\sigma_i,\sigma_{\partial i}]  \Big[\prod_{k\in\partial i \setminus j } p(\sigma_k|\s_i)\Big] p(\s_i|\s_j) \\
 &-  r_{i}[-\sigma_i,\sigma_{\partial i}] \Big[\prod_{k\in\partial i \setminus j } p(\sigma_k|-\s_i)\Big] p(-\s_i|\s_j) \Bigg]
\label{eqn:alltogether3_approx}
\end{align}

\noindent that we call {\em Cavity Master Equation} for reasons that will be clear below. 

Equation \eqref{eqn:alltogether3_approx} is the main result of our work and together with \eqref{eq:localMEfact} provides a closed set of equations for the dynamics of the single site. In Section \ref{sec:ExactResults} we show that this equation reproduces analytical exact results for both the mean-field and the  one-dimensional Ising ferromagnet. In Section \ref{sec:num_results} we test the performances of the closure scheme \eqref{eq:localMEfact} and \eqref{eqn:alltogether3_approx} for different models defined on random graphs. The reader not interested in the analytical derivation of \eqref{eq:localMEfact} can skip the following sections and continue from Sec. \ref{sec:num_results}.

\section{Random Point Processes}\label{sec:RPP}

In this section we introduce the Random Point Process formalism which will be used to parametrized probability distributions of spin histories in continuous time. To get familiar with the notation we first concentrate on just one independent spin. 

 The probability of having a single spin in state $\sigma$ at time $t$ given the initial condition $\sigma_0$ at time $t_0$, $p(\sigma,t|\sigma_0,t_0)$, can be specified by the sum of the probability weight of all trajectories that transform this initial state $\sigma_0$ into $\sigma$ after a time $t-t_0$. A specific spin history or trajectory $X$ is parametrized by the number of spin flips, the time in which they occur and the initial state of the system. The spin trajectory is then nothing but a Random Point Process (RPP) \cite{van1992stochastic} and the probability measure in this sample space may be denoted as
\begin{equation}
Q(X) =  Q_s(t_0,t_1,\dots t_s,t| \sigma(t_0)=\sigma_0)
\label{eq:RPP}
\end{equation}
\noindent which represents the probability density of having a trajectory with $s$ jumps at $(t_1,t_1+dt_1)\ldots(t_s,t_s+dt_s)$ etc. given the initial state $\s_0$. We here stress a detail on notation: there is a difference between $\s(\tau)$, which is the value of the spin orientation at the particular time $\tau$ and the complete trajectory $X$ of such spin, which is specified by the value that $\s(\tau)$ takes for every time in the interval $[t_0,t]$. When needed,
we may write $X$ as $X(t)$ to emphasize that the final time of the spin history is precisely $t$.

To recover $p(\sigma,t|\sigma_0,t_0)$ from $Q(X)$, one integrates $Q_s(t_0,t_1,\dots t_s,t| \sigma(t_0)=\sigma_0)$ over all times for a fixed $s$ and sums over all possible values of $s$. For example, in order to find $p(\sigma=\sigma_0,t|\sigma_0,t_0)$, the probability of having the same
orientation at time $t$ as in the initial state, we have to sum over all $s=2k$ possible values:
\begin{equation}
 p(\s_0,t|\s_0,t_0) = Q_0 + \sum_{k=1}^{\infty}\int_{t_0}^{t}dt_1\int_{t_1}^{t}dt_2\ldots\int_{t_{s-1}}^{t}dt_s Q_{2k}(t_0,...,t)
 \label{eqn:marginalizing_Q}
\end{equation}
In a nutshell, we sum the probability density of all trajectories with an even number of jumps because those, and only those, transform 
the initial state into itself.

If the spin flips occur independently at a rate that depends only on the instantaneous orientation $\s$ and not on previous values, 
the inter-jump waiting time is exponentially distributed and
\begin{equation}
 Q_{s=2k}=r(\sigma_0) e^{-r(\sigma_0)(t_1-t_0)} r(-\sigma_0)e^{-r(-\sigma_0)(t_2-t_1)} \ldots r(-\sigma_0)e^{-r(-\sigma_0)(t_{s}-t_{s-1})}e^{-r(\sigma_0)(t_{f}-t_{s})},
\label{eqn:density_random_proc_explicit} 
\end{equation}
\noindent where $r(\s)$ is the jumping rate that, in this case of one-single spin, only depends on the spin configuration $\sigma$ at time $t$. 

If instead of a single spin we deal with the more general case of $N$ interacting spins, the set of the individual histories $X_1,\ldots X_N$ can still be parametrized by the number of jumps of the corresponding spin $s_i$ and the time coordinate of each trajectory, so
$ X_i  \sim \{ s_i, (t_1^i,t_2^i,\ldots,t_{s_i}^i) \} \sim \{ s_i, \vec{ t}\,^i \}$. For a tree-like topology, the full (density) probability distribution of all the histories in the system reads as
\begin{equation}
 Q(X_1,\ldots,X_N) = \prod_{a=1}^N \left \lbrace  \prod_{l_a=1}^{s_a} r_a\left[\s_a(t_{l_a}), \s_{\partial a}(t_{l_a})\right]
                    e^{-\int_{t_0}^t r_a\left[\s_a(\tau), \s_{\partial a}(\tau)\right]) d\tau } \right\rbrace
\label{eqn:full_joint_prob_dist}
\end{equation}
\noindent where $r_a$ is the jumping rate of spin $a$, given the state of its neighbours in $\partial a$, similar to the transition rate appearing in the master equation \eqref{eq:localME}.
We can shortly write the quantity inside the curly brakets as $\Phi_a(X_a|X_{\partial a})$, since it represents the probability density of the history $X_a$ with the histories of the neighbourhood $X_{\partial a}$ fixed. So, equation (\ref{eqn:full_joint_prob_dist}) can be rewritten in a compact form as
\begin{equation}
  Q(X_1,\ldots,X_N) = \prod_{a=1}^N \Phi_a(X_a|X_{\partial a})
  \label{eqn:full_joint_prob_dist2}
\end{equation}

\section{Dynamic Message Passing Equations}


In this section, starting from $Q(X_1,\ldots,X_N)$, we re-derive the dynamic message-passing (or dynamic cavity equation) as presented in \cite{del2015dynamic} for the case of discrete-time update dynamics. This equation is an iterative relation for conditional probabilities of spin histories, exact on tree-like graph in the thermodynamic limit. Then by using the RPP formalism we parametrize these probabilities and we give a proper mathematical description of how to marginalize them over time for the case of continuous-time processes. The combined use of the dynamic message-passing equation and the RPP formalism is our starting point to obtain equation \eqref{eqn:alltogether3_approx}.

Assuming a tree-like architecture network, we start selecting a spin, say $i$, and rewrite equation (\ref{eqn:full_joint_prob_dist2}) expanding the tree around it and making use of its structure:
\begin{equation}
    Q(X_1,\ldots,X_N) = \Phi_i(X_i|X_{\partial i}) \prod_{k \in \partial i} \left[ \Phi_k(X_k|X_{\partial k})  
                                                   \prod_{m \in \partial k \setminus i} \left[ \Phi_k(X_m|X_{\partial m}) \prod_{l \in \partial m \setminus k} \ldots  \right]\right]
                                                   \label{eqn:full_joint_prob_dist_tree}
\end{equation}
Let $G_k^{(i)}$ be the sub-graph expanded from the site $k$ after removing the link $(ik)$. We define as $\{X\}_{ik}$ the set of histories of the spins included in $G_k^{(i)}$ except
$X_k$ itself. With these definitions we express (\ref{eqn:full_joint_prob_dist_tree}) as:
\begin{equation}
    Q(X_1,\ldots,X_N) = \Phi_i(X_i|X_{\partial i}) \prod_{k \in \partial i} M_{ki}(X_k,\{X\}_{ik}|X_i)
\end{equation}
Here $M_{ki}$ is just a shorthand for the expression inside brackets. Marginalizing $Q$ on all histories except $X_i,X_{\partial i}$ we get the joint (density) probability distribution of the history of spin $i$ and its neighbours
\begin{equation}
    Q(X_i,X_{\partial i}) = \Phi_i(X_i|X_{\partial i}) \prod_{k \in \partial i} \mu_{k\rightarrow (ki)}(X_k|X_i),
     \label{eqn:marginal_prob_dist_mess}
\end{equation}
\noindent where the new functions $\mu_{k\rightarrow (ki)}(X_k|X_i)$ are the marginal of $M_{ki}(X_k,\{X\}_{ik}|X_i)$:
\begin{equation}\label{eq:mu}
   \mu_{k\rightarrow (ik)}(X_k|X_i) = \sum_{\{X\}_{ik}} M_{ki}(X_k,\{X\}_{ik}|X_i)
\end{equation}

\noindent and have clearly the interpretation of the probability of history $X_k$ given $X_i$ fixed. 
If we take now two neighbours, $i$ and $j$, and make a similar reasoning we conclude that their marginal probability can be written as:
\begin{equation}
    Q(X_i,X_j) =  \mu_{i\rightarrow (ij)}(X_i|X_j) \mu_{j\rightarrow (ji)}(X_j|X_i)
    \label{eqn:pair_prob_dist_mess}
\end{equation}
The final step in order to derive the Dynamic Message-Passing equation is to marginalize (\ref{eqn:marginal_prob_dist_mess}) on $X_{\partial i \setminus \{i,j\}}$ and combine with (\ref{eqn:pair_prob_dist_mess}). Simplifying terms, we get 
\begin{equation}
 \musubX{i}{j} = \sum_{\{X_k\},k\in\partial i\setminus j} \Phi_i (X_i|X_{\partial i}) \prod_{k\in\partial i\setminus j} \musubX{k}{i}
 \label{eqn:update_bethe}
\end{equation}
\noindent where $X_i = X_i(t)$ is the history of spin $i$ up to time $t$ and the trace is over all the spin trajectories for $k \, \in \p i \backslash \, j$. To simplify notation we will sometimes write $\musubsim{i}{j}{t}$ for the cavity conditional probability and, according to the previous literature \cite{YedidiaFreemanWeiss2003,mezard2009information,del2015dynamic}, we may refer to it as ``message''.

The probabilistic content of $\mu(X_i|X_j)$ is subtle. Note that following the usual convention for defining conditional probabilities we might write
\begin{equation}
 Q(X_i|X_j) =  \dfrac{Q(X_i,X_j)}{Q(X_j)}
 \label{eqn:Qconditional}
\end{equation}
therefore it is natural to ask what is the connection of this quantity to the cavity conditional probability. The key to understand the difference lies
precisely on the way they were defined. The cavity message $\mu(X_i|X_j)$ is the marginal probability of $X_i$ in a graph where $X_j$ is a boundary condition. In 
other words, it is the distribution of $X_i$ for a \textit{fixed} $X_j$. On the other hand, $Q(X_i|X_j)$ is the probability of having $X_i$ if we know
that $X_j$ occurred\footnote{We use a simplified language but it should be clear that we are dealing with probability densities.}. Roughly speaking,
we measure $\mu(X_i|X_j)$ by counting how many times $X_i$ appears when $X_j$ is fixed, while for $Q(X_i|X_j)$ we should pick from all the runs where $X_j$ is realized, those where we find $X_i$. The mathematical difference between $\mu(X_i|X_j)$ and $Q(X_i|X_j)$ will be further discussed at the end of Section \ref{sec:CME}. We also refer to \cite{aurell2016causal} where similar probabilistic measures are discussed and compared.

Finally, let us comment that for the discrete-time case, the meaning of the marginalization over a spin trajectory as one of those appearing on the RHS of \eqref{eqn:update_bethe} is clear: it is indeed a sum over all the possible configurations of the spin at discrete time steps $\{\s_k(0),\dots,\s_k(t-1),\s_k(t)\}$. For the continous-time case, a spin trajectory is parametrized according to the RPP formalism as described in Section \ref{sec:RPP}, \emph{i.e.} $ X_k  \sim \{ s_k, (t_1^k,t_2^k,\ldots,t_{s_k}^k) \}$. Therefore spin trajectory marginalizations, as the traces above, in the continuous time case must be interpreted as 
\begin{equation}
\sum_{X_k} \,(\cdot) = \sum_{s_k} \left[ \int_{t_0}^{t}dt_1^{k}\int_{t_1^{k}}^{t}dt_2^{k}\ldots \int_{t_{s_k-1}^{k}}^{t}dt_{s_k}^{k} \,(\cdot) \right]
\label{eq:traces}
\end{equation}
In the following sections we will use this mathematical interpretation given by RPP formalism for the derivation of the Cavity Master Equation \eqref{eqn:alltogether3_approx}. 

\subsection{Differential update equations}

We have already shown in \eqref{eqn:full_joint_prob_dist} that probability densities, within the RPP formulation, can be parametrized as
\begin{equation}
 \musub{i}{j}{t} = \prod_{l_i=1}^{s_i} \lasubsim{i}{j}{t_{l_i}} \exp\{ - \int_{t_0}^{t} \lasubsim{i}{j}{\tau} d\tau \} 
 \label{eqn:cavity_mess_param}
\end{equation}
where above $t_{l_i}$ are the jumping times for the $X_i$ history, which has $s_i$ jumps. In this equation, $\lambda_{i\rightarrow (ij)}$ is interpreted as the effective jumping rate of $i$ at each time, with a given history of $j$. A way to picture
the meaning of $\mu_{i \rightarrow (ij)}$ would be imagining a spin $i$ flipping under the influence of a external field $h(\tau)$. This external field will be the sum of two effects: the explicit interaction with spin $j$ and a term for the average interaction with the other neighbours
\begin{equation}
 h(\tau) = J_{ij} \s_j(\tau) + h_{\partial i \setminus j}(\tau)
\end{equation}
Then $\lasubsim{i}{j}{\tau}$ will just be the jumping rate of spin $i$ under $h(\tau)$. The average interaction with neighbours different from $j$ is fixed if the previous history of $X_i$ is given, since $X_i$ is a boundary condition for the evolution of the tree starting from $i$ and growing in the direction opposite to $j$. In the more general case then, $\lambda_{i \to (ij)}$ is a function of the spin history of the variable $i$ and $j$ and thus, for a spin dynamics up to time $\tau$ we may more explicitly write $\lasubsim{i}{j}{\tau} = \lasubsim{i}{j}{X_i,X_j, \tau}$.

On the other hand, the interaction term $\Phi_i (X_i(t)|X_{\partial i}(t))$ in \eqref{eqn:update_bethe}, already introduced through \eqref{eqn:full_joint_prob_dist} and \eqref{eqn:full_joint_prob_dist2}, can be interpreted as the probability density of $X_i$ conditioned on the histories of spins in $\partial i$:
\begin{equation}
 \Phi_i (X_i(t)|X_{\partial i}(t)) = \prod_{l_i=1}^{s_i} r_{i}(\sigma_i(t_{l_i}),\sigma_{\partial i}(t_{l_i}))  
                                                         \exp\{ - \int_{t_0}^{t} r_{i}(\sigma_i(\tau),\sigma_{\partial i}(\tau)) d\tau \}
\end{equation}
\noindent where $r_i$ is the jumping rate of spin $i$. For a Markov dynamics this is an instantaneous quantity, meaning that at time $\tau$ it depends only on the values of $\sigma_{i}(\tau)$ and  $\sigma_{\partial i}(\tau)$. 

According to \eqref{eq:traces}, the trace on the right hand side of \eqref{eqn:update_bethe} can be written in more detail. By calling $F$ the argument of the sum, we can rewrite the RHS as 

\begin{equation}
 \sum_{\{X_k\},k\in\partial i\setminus j}^{[t_0,t]} F(X_i,X_{\partial i},t) = \sum_{\{s_k\},k\in\partial i\setminus j} 
								  \left[ \prod_{k=1}^{d} \int_{t_0}^{t}dt_1^{k}\int_{t_1^{k}}^{t}dt_2^{k}\ldots \int_{t_{s_k-1}^{k}}^{t}dt_{s_k}^{k}\right]
								  F(X_i,X_{\partial i},t)
\end{equation}
%
\newline
We here stress that, as for the LHS of \eqref{eqn:update_bethe}, the function F above is a conditional probability although, to simplify notation, we do not highlight it explicitly. 

In principle, through the parametrization (\ref{eqn:cavity_mess_param}), if we write (\ref{eqn:update_bethe}) for every pair $(i,j)$ in the network we get a system of coupled equations for the $\lambda$'s. Solving them we may describe the dynamics of the system. Unfortunately (\ref{eqn:update_bethe}) is a very involved expression that needs to be transformed in a more tractable one. We here propose to differentiate it with respect to the parameter $t$, the final time, in order to have a differential equation for the messages. 

Differentiation in this context should be handled carefully since increasing $t$ means we are changing the sample space itself. Writing  $\musub{i}{j}{t+\Delta t}$ in terms of $\musub{i}{j}{t}$ is mapping the probability of one sample space (that of all possible histories up to time $t$) onto another (histories up to time $t+\Delta t$). Instead of using standard differentiation rules it is safer to go by the definition.   
Therefore, for the left hand side of equation (\ref{eqn:update_bethe}) we will compute the limit:

\begin{equation}
 \lim_{\Delta t\rightarrow 0} \dfrac{\musub{i}{j}{t+\Delta t} - \musub{i}{j}{t}}{\Delta t}
\end{equation}
 A very important question arises at this point. What is the relation of $(X_i(t+\Delta t),X_j(t+\Delta t))$ and $(X_i(t),X_j(t))$? Or in other
 words, what happens in the interval $(t,t+\Delta t)$? The answer is important because expressions for $\musub{i}{j}{t+\Delta t}$ are different
 whether we consider that there can be jumps in the small $\Delta t$ interval or not. The first thing that makes sense to impose is that histories
 must agree up to time $t$. In $(t,t+\Delta t)$ we can have several combinations.
 
 An implicit assumption throughout all this theory is that on an infinitesimal interval only two things can happen to a spin; it can
 stay on its current state or make one and only one jump to the opposite orientation. Two or more jumps are not allowed. Considering this we have four 
 cases to analyze:
 \begin{itemize}
  \item There are $s_i,s_j$ jumps in $(t_0,t)$ and neither $i$ nor $j$ jumps in $(t,t + \Delta t)$. This occurs with a probability
   $(1-\lambda_i \Delta t)(1-\lambda_j \Delta t)$.
  \item There are $s_i,s_j$ jumps in $(t_0,t)$ and $i$ XOR $j$ jumps in $(t,t + \Delta t)$. This occurs with a probability
   $(1-\lambda_i \Delta t)(\lambda_j \Delta t)$ or  $(1-\lambda_j \Delta t)(\lambda_i \Delta t)$. These are two cases in one.
  \item There are $s_i,s_j$ jumps in $(t_0,t)$ and both $i$ and $j$ jumps in $(t,t + \Delta t)$. This has a probability of $\lambda_j \lambda_i \Delta t ^2$
 \end{itemize}
 When $\Delta t$ goes to zero, from the previous analysis we conclude that the derivative should be computed, with probability 1, using the first option, where histories for $i$ and
 $j$ have no jumps in the interval of length $\Delta t$.
 
To differentiate the LHS of (\ref{eqn:update_bethe}) we use the parametrization (\ref{eqn:cavity_mess_param}). Writing shortly $\mu_{i \to (ij)}(t)$ for the message $ \musub{i}{j}{t}$ at time $t$, we get
\begin{eqnarray}
\nonumber
\mu_{i \to (ij)}(t+ \Delta t) &=& \prod_{l_i=1}^{s_i} \lasubsim{i}{j}{t_{l_i}} \exp\{ - \int_{t_0}^{t + \Delta t} \lasubsim{i}{j}{\tau} d\tau \} \\
 \nonumber
 &\approx& [1-\lasubsim{i}{j}{t} \Delta t] \prod_{l_i=1}^{s_i} \lasubsim{i}{j}{t_{l_i}} \exp\{ - \int_{t_0}^{t} \lasubsim{i}{j}{\tau} d\tau \} + o(\Delta t) \\
 &=& [1-\lasubsim{i}{j}{t} \Delta t]\:\mu_{i \to (ij)}(t) + o(\Delta t) 
\end{eqnarray}
 Then, with probability 1, the derivative of the LHS of \eqref{eqn:update_bethe} is equal to $ -\lasubsim{i}{j}{t}\:\mu_{i \to (ij)}(t)$.
 
 To obtain the derivative of the RHS of equation \eqref{eqn:update_bethe}, we must compute:
 \begin{equation}
 \lim_{\Delta t \rightarrow 0} \dfrac{\displaystyle \sum_{\{X_k\},k\in\partial i\setminus j}^{[t_0,t + \Delta t]} F(X_i,X_{\partial i},t + \Delta t) - \musub{i}{j}{t} }{\Delta t}
 \end{equation}
\
Let us focus on the first term in the numerator of the previous expression. It can be expanded to first order in $\Delta t$. It is important to remember that $\Delta t$ appears
in the integration limits as well as in the integrand of $F$. In addition, we should keep in mind that all jumps for $X_i$ and $X_j$ must occur before $t$. This restriction, though, does not
apply to the histories $X_k $ for $k$ in $\partial i \setminus j$.

The expansion of \eqref{eq:F} is a lenght calculation. However, we can paraphrase the idea in a few lines. First, let us remember that $F$ is the joint probability of $X_i$ and $\{X_k\}$ with $k\in \partial i \setminus j$, conditioned on $X_j$. All the histories of interest are in the interval $[t_0,t+\Delta t]$.
The expression:
\begin{equation}
  \sum_{\{X_k\},k\in\partial i\setminus j}^{[t_0,t + \Delta t]} F(X_i,X_{\partial i},t + \Delta t)
 \label{eq:F}
\end{equation}
\noindent is the marginalization of the mentioned joint probability distribution. The previous sum, to order $\Delta t$, has two main contributions. One comes from summing over $\{X_k\}$ with all $X_k$ having no jumps in $[t,t+\Delta t]$:
\begin{equation}
 A =  \sum_{\{X_k\},\, k\in\partial i\setminus j}^{[t_0,t]} F(X_i,X_{\partial i},t) \Big[1 - \big(\sum_k \lasubsim{k}{i}{t} + r_i(t)\big)\Delta t\Big]
 \label{eq:A}
\end{equation}
The other considers all the possibilities of having one of the $X_k$ with a jump in the interval of
length $\Delta t$:
\begin{equation}
 B =\displaystyle \sum_{k} \sum_{\{X_k\},\, k\in\partial i\setminus j}^{[t_0,t]} F(X_i,X_{\partial i},t)  \lasubsim{k}{i}{t}\Delta t
\end{equation}
The expansion to first order in $\Delta t$ of equation \eqref{eq:F} then reads
\begin{equation}
 \sum_{\{X_k\},k\in\partial i\setminus j}^{[t_0,t + \Delta t]} F(X_i,X_{\partial i},t + \Delta t) = A + B + o(\Delta t)
\end{equation}
Writing the right hand side of the above equation explicitly we observe that $B$ cancels out with the $\lambda$ part of $A$ in \eqref{eq:A} and the remaining term of order 1 is $\musub{i}{j}{t}$, which
cancels when inserted in the limit expression. Then, the final outcome of the differentiation of equation \eqref{eqn:update_bethe} reads
\begin{equation}
 \lasubsim{i}{j}{X_i,X_j, t}\,\musubX{i}{j} = \displaystyle \sum_{\{X_k\},k\in\partial i\setminus j}^{[t_0,t]} 
 r_{i}[\sigma_i(t),\sigma_j(t),\sigma_{\partial i \setminus j}(t)] F(X_i,X_{\partial i},t)
\end{equation}
\noindent We can now marginalize the right hand side of the above equation of all the spin histories of the spins $k \, \in \p i \backslash \, j$ by keeping the configuration of these spins at the last time $t$ fixed. The results reads
\begin{equation}
\lasubsim{i}{j}{X_i,X_j, t}\, \musubX{i}{j} = \displaystyle \sum_{\sigma_{\partial i\setminus j}(t)} r_{i}[\sigma_i(t),\sigma_j(t),\sigma_{\partial i \setminus j}(t)] P(\sigma_{\partial i \setminus j}(t),X_i|X_j)
 \label{eqn:lambda_mu}
\end{equation}
\noindent where we introduced the function $P$ as the marginalization of the function $F$ over all the spin histories of the $i$-neighbours except $j$, with the configuration at the final time fixed. Note that the notation $\sigma_{\partial i\setminus j}(t)$ is equivalent to $\{\s_{k}(t)\}_{k \in \p i \backslash j}$ and that in $P$ above appears again explicitly the conditional nature of the probability distribution $F$. 

Equation \eqref{eqn:lambda_mu} represents the differential dynamic-message passing update equation obtained by differentiating \eqref{eqn:update_bethe} in time. It connects the derivative of the dynamic message $\mu_{i \to (ij)}$, and so the effective jumping rate $\lambda_{i \to (ij)}$ of spin $i$ used to parametrize the message in \eqref{eqn:cavity_mess_param}, with the transition rate of the same spin $r_{i}[\sigma_i(t),\sigma_j(t),\sigma_{\partial i \setminus j}(t)]$ at time $t$. This result will be used in next section for our final derivation.

\section{The cavity master equation}\label{sec:CME}

The cavity messages $\musubX{i}{j}$ is a complicated object with high dimensionality. It is a real valued \textit{functional} of $X_i$ given the history $X_j$, where both $\mu$ and $X$ depend parametrically on $t$. For our purposes, it is convenient to reduce the dimensionality of this message by partially marginalizing over the spin history of spin $i$. We so introduce an easier mathematical object to deal with, which is the marginal of $\musubX{i}{j}$ with the final state fixed
\begin{equation}
 p(\s_i,t|X_j) = \sum_{X_i | \s_i(t) = \s_i} \musubX{i}{j}
 \label{eqn:conditional_prob_single_spin_on_Xj}
\end{equation}
By differentiating the above equation we can obtain an equation for the evolution of this probability distribution. As we have seen in the previous section, derivative must be computed by using the standard definition of the limit of the increment ratio:
\begin{equation}
 \dot{p}(\s_i,t|X_j) = \lim_{\Delta t \rightarrow 0} \dfrac{p(\s_i,t + \Delta t|X_j(t + \Delta t)) - p(\s_i,t|X_j(t + \Delta t))}{\Delta t}
 \label{eqn:increment_ratio}
\end{equation}
 We hereafter write $p(\s_i,t+\Delta t| X_j(t+\Delta t))$ as the marginalization of  $\mu_{i \rightarrow (ij)}(X_i(t+\Delta t)|X_j(t+\Delta t))$ and then, following the procedure developed in the last section, we expand it to first order in $\Delta t$, similarly to what we did for \eqref{eq:F}. Using the usual short notation for $\mu$, we get explicitly: 
\begin{equation}
 p(\s_i,t+ \Delta t|X_j(t+\Delta t)) = \sum_{s=0}^{\infty}\int_{t_0}^{t + \Delta t}dt_1\int_{t_1}^{t + \Delta t}dt_2\ldots\int_{t_{s-1}}^{t + \Delta t}dt_s \, 
\mu_{i \rightarrow (ij)}(t+\Delta t)
 \label{eqn:marginalizing_Q_general_delta_t}
\end{equation}
Let us call the series of the $s$ integrals above as $I_{t + \Delta t}$ and note that they can be written separating the $o(\Delta t)$ terms as follows
\begin{equation}
 I_{t + \Delta t}  \overset{.}{=} \int_{t_0}^{t + \Delta t}dt_1\int_{t_1}^{t + \Delta t}dt_2\ldots\int_{t_{s-1}}^{t + \Delta t}dt_s 
= \int_{t_0}^{t }dt_1\int_{t_1}^{t}dt_2\ldots\int_{t_{s-1}}^{t + \Delta t}dt_s + o(\Delta t)
 \label{eqn:integral_expansion}
\end{equation}
and that, furthermore, the last one of them can be split into two intervals from $[t_s,t]$ and $[t,t+\Delta t]$
\begin{equation}
I_{t + \Delta t} = \int_{t_0}^{t }dt_1\int_{t_1}^{t}dt_2\ldots\int_{t_{s-1}}^{t}dt_s \;+ \; \int_{t_0}^{t }dt_1\int_{t_1}^{t}dt_2\ldots\int_{t}^{t + \Delta t}dt_s + o(\Delta t)\label{eqn:integral_expansion_1}
\end{equation}
The first term in (\ref{eqn:integral_expansion_1}) is a trace over trajectories that have all jumps in $[t_0,t]$; the second term considers
that the $s$-th jump occurs in $[t,t+\Delta t]$. We then insert $\mu_{i \rightarrow (i,j)}(t+\Delta t)$ into $I_{t + \Delta t}$ expressed as above in order to compute \eqref{eqn:marginalizing_Q_general_delta_t}. Let us observe that in the first integral $\mu_{i \to (ij)}(t + \Delta t)$ should be expanded to first order in $\Delta t$, whereas in the second integral it can be left to the order zero expansion since the integral has an order $\Delta t$ itself.

With the usual compact notation, the expansion of $\mu_{i \rightarrow (i,j)}(X_i(t+\Delta t)|X_j(t+\Delta t))$ reads:
\begin{eqnarray}
 \mu_{i \rightarrow (i,j)}(t+\Delta t) &=& \Big[ \prod_{i=1}^{s} \lambda(t_i) \Big] e^{-\int_{t_0}^{t + \Delta t} \lambda(\tau) d\tau} = \mu_{i \rightarrow (i,j)}(t) \left[1 - \lambda (t) \Delta t + o(\Delta t) \right]
\end{eqnarray}
\noindent
where hereafter we omit the subscript of $\lambda$ to shorten notation. Adding all together these results  as described above and collecting $o(\Delta t)$ terms we obtain
\begin{eqnarray}
\nonumber
 \lefteqn{p(\s_i,t+ \Delta t|X_j(t+\Delta t)) =  \sum_{X_i | \s_i(t) = \s_i}\left[1 - \lambda (t) \Delta t \right]  \mu_{i \rightarrow (i,j)}(X_i(t)|X_j(t))}\\
&& + \sum_s \int_{t_0}^{t }dt_1\int_{t_1}^{t}dt_2\ldots\int_{t}^{t + \Delta t}dt_s  \left[ \prod_{i=1}^{s} 
\lambda(t_i) \right] e^{-\int_{t_0}^{t} \lambda(\tau) d\tau}+ o(\Delta t)
\label{eqn:iterated_integral_approximation}
\end{eqnarray}
As previously mentioned, the last integral on the second term of (\ref{eqn:iterated_integral_approximation}) corresponds to the probability of having the last jump $s$ in the interval $[t, t + \Delta t]$. 
Since the orientation before the last jump is $-\s_i$,
the corresponding jumping rate in this case is $\lambda(t) = \lambda(-\s_i(t),X_i^-,X_j)$, where to stress this difference we separated the value of the last spin of $i$ at time $t$ from its previous history and $X_i^{-}$ denotes that the final state of this history is $-\s_i$. With this notation, last integral in \eqref{eqn:iterated_integral_approximation} can be expanded as
\vspace{0.1cm}
\begin{eqnarray}
\nonumber
\lefteqn{\int_{t_0}^{t }dt_1\int_{t_1}^{t}dt_2\ldots\int_{t}^{t + \Delta t}dt_s  \left[ \prod_{i=1}^{s} 
\lambda(t_i) \right] e^{-\int_{t_0}^{t} \lambda(\tau) d\tau}=}\\
\nonumber
&&=\int_{t_0}^{t }dt_1\int_{t_1}^{t}dt_2\ldots\int_{t_{s-2}}^{t}dt_{s-1}
\left\lbrace  \left[ \prod_{i=1}^{s-1} \lambda(t_i) \right] e^{-\int_{t_0}^{t} \lambda(\tau) d\tau}  \lambda(-\s_i(t),X_i^-, X_j)\right\rbrace \Delta t 
+ o(\Delta t)\\
&&=\int_{t_0}^{t }dt_1\int_{t_1}^{t}dt_2\ldots\int_{t_{s-2}}^{t}dt_{s-1}
\left\lbrace \mu_{i \rightarrow (i,j)}(X_i^-(t)|X_j(t))  \lambda(-\s_i(t),X_i^-,X_j) \right\rbrace \Delta t 
+ o(\Delta t)
\label{eqn:iterated_integral_approximation2}
\end{eqnarray}

\noindent We highlight that the histories $X_i^{-}$ and $X_j$ as arguments of $\lambda$ above run up to time $t$ and, as mentioned, the superscript in $X_i^-$ is included to explicitly state that the corresponding integrals are tracing histories of $i$ with $s-1$ jumps and it means that
the last state is the opposite to $\sigma_i(t)$.

Combining together the results in \eqref{eqn:marginalizing_Q_general_delta_t}, \eqref{eqn:iterated_integral_approximation} and \eqref{eqn:iterated_integral_approximation2}  
and taking the limit $\Delta t \rightarrow 0$ we get that the derivative expressed in \eqref{eqn:increment_ratio} reads
\begin{eqnarray}
\label{eqn:master_equation_continuous_time}
 \dot{p}(\s_i,t|X_j) &=& - \sum_{X_i | \s_i(t) = \s_i} \lambda_{i \rightarrow (ij)} [\sigma_i(t),X_i,X_j] \;  \musubX{i}{j}\\
 &&+  \sum_{X_i^- | \s_i(t) = -\s_i}  \lambda_{i \rightarrow (ij)} [-\sigma_i(t),X_i^-,X_j] \; \mu_{i \rightarrow (i j)}(X_{i}^-|X_{j})
\nonumber
 \end{eqnarray}
%
Finally, plugging (\ref{eqn:lambda_mu}) into (\ref{eqn:master_equation_continuous_time}) and after some re-arrangement:
 \begin{align}
 \nonumber
 \dot{p}(\s_i,t|X_j) = -   \sum_{\sigma_{\partial i\setminus j}} \Big[ &\sum_{X_i | \s_i(t) = \s_i} r_{i}[\sigma_i,\sigma_{\partial i}] p(\sigma_{\partial i \setminus j},X_i|X_j)\\
 \label{eqn:alltogether1}
 &-  \sum_{X_i^- | \s_i(t) = -\s_i} r_{i}[-\sigma_i,\sigma_{\partial i}] p(\sigma_{\partial i \setminus j},X_i^-|X_j)\Big]\\
 \nonumber
 =& -  \sum_{\sigma_{\partial i\setminus j}} \Big[r_{i}[\sigma_i,\sigma_{\partial i}]  p(\sigma_{\partial i \setminus j}, \s_i|X_j)\\
 &\hspace{1.3cm} -  r_{i}[-\sigma_i,\sigma_{\partial i}] p(\sigma_{\partial i \setminus j},-\s_i|X_j)\Big]
\label{eqn:alltogether2}
\end{align}

\noindent where in the second equality we performed the traces inside the brackets. 

Let us emphasize that to arrive at the result expressed in Eq. \eqref{eqn:alltogether2} we did not take any specific approximation but only assumed that the graph topology is tree-like. To continue however, backed by the Markov character of the model, we may assume that the probability distribution of an instantaneous variable, conditioned on the history of a neighbour depends only on the last state of that neighbour: $p(\sigma_i|X_j) \approx p(\sigma_i|\sigma_j)$. In this case, (\ref{eqn:alltogether2}) transforms into:
\begin{align}
 \dot{p}(\s_i,t|\s_j) = - \displaystyle \sum_{\sigma_{\partial i\setminus j}}\Big[ r_{i}[\sigma_i,\sigma_{\partial i}]  \,p(\sigma_{\partial i \setminus j}, \s_i|\s_j) - r_{i}[-\sigma_i,\sigma_{\partial i}]\, p(\sigma_{\partial i \setminus j},-\s_i|\s_j) \Big]
\label{eqn:alltogether2_approx}
\end{align}
Then, in order to close the equations we assume that the conditional distributions factorize as follows
\begin{eqnarray}\nonumber
 p(\sigma_{\partial i \setminus j}, \s_i|\s_j) &=& p(\sigma_{\partial i \setminus j}|\s_i,\s_j) p(\s_i|\s_j)\\
 &\approx& p(\sigma_{\partial i \setminus j}|\s_i) p(\s_i|\s_j) \approx \Big[\prod_{k\in\partial i \setminus j } p(\sigma_k|\s_i)\Big] p(\s_i|\s_j)
 \label{eqn:approx_prob}
\end{eqnarray}
Plugging equation \eqref{eqn:approx_prob} into \eqref{eqn:alltogether2_approx} lead to the {\em Cavity Master Equation}, already anticipated in \eqref{eqn:alltogether3_approx}:
\begin{align}
 \nonumber
 \frac{d p(\s_i|\s_j)}{d t} = -  \sum_{\sigma_{\partial i\setminus j}} \Bigg[& r_{i}[\sigma_i,\sigma_{\partial i}]  \Big[\prod_{k\in\partial i \setminus j } p(\sigma_k|\s_i)\Big] p(\s_i|\s_j) \\
 &-  r_{i}[-\sigma_i,\sigma_{\partial i}] \Big[\prod_{k\in\partial i \setminus j } p(\sigma_k|-\s_i)\Big] p(-\s_i|\s_j) \Bigg]
\label{eq:CME}
\end{align}
Before ending this section we should discuss the relation linking the cavity conditional probability $p(\s_i|\s_j)$ with the one appearing in
(\ref{eq:localMEfact}), $P(\s_i|\s_k)$. We can start marginalizing the pair distribution:
\begin{equation}
 P(\s_i|X_j) =\sum_{X_i | \s_i(t) = \s_i} Q(X_i|X_j) = \sum_{X_i | \s_i(t) = \s_i} \dfrac{Q(X_i,X_j)}{Q(X_j)}
\end{equation}
which, if we use (\ref{eqn:pair_prob_dist_mess}) becomes:
\begin{eqnarray}
\nonumber
 P(\s_i|X_j) &=& \sum_{X_i | \s_i(t) = \s_i} \mu_{i\rightarrow (ij)}(X_i|X_j)  \dfrac{\mu_{j\rightarrow (ji)}(X_j|X_i)}{\sum_{X_i} \mu_{i\rightarrow (ij)}(X_i|X_j) \mu_{j\rightarrow (ji)}(X_j|X_i)}\\
&=& \sum_{X_i | \s_i(t) = \s_i} \mu_{i\rightarrow (ij)}(X_i|X_j) \Delta \mu_{j\rightarrow (ji)}(X_j,X_i)
 \label{eqn:Pexpanded}
\end{eqnarray}
where in the second equality we introduced the quantity $ \Delta \mu_{j\rightarrow (ji)}(X_j,X_i)$ as the rate appearing on the RHS of the first equality. It is easy to show that in a low correlation regime $ \lim_{c_{ij}\rightarrow 0} \Delta \mu_{j\rightarrow (ji)}(X_j,X_i)  \simeq 1$ (with $c_{ij}$ the correlation between site $i$ and $j$). Therefore,
thanks to (\ref{eqn:Pexpanded}) we can state that $p(\s_i|X_j)$, as defined in (\ref{eqn:conditional_prob_single_spin_on_Xj}), corresponds in this limit to $P(\s_i|X_j)$:

\begin{equation}
 P(\s_i|X_j) = p(\s_i|X_j) + g(c_{ij}) \;\;\; \mbox{where}\;\;\; \lim_{c_{ij}\rightarrow 0} g(c_{ij}) = 0
\end{equation}

Moreover, if correlations are weak, it should also be the case that $P(\s_i|X_j) \approx P(\s_i|\s_j) = p(\s_i|\s_j)$. This is the reason why we plug the results found in (\ref{eq:CME}) into (\ref{eq:localMEfact}) to describe the dynamics of the
system. From a practical point of view (\ref{eq:CME}), is the main result of this work.

\section{Comparison with known exact results}\label{sec:ExactResults}

In this section we show that the Cavity Master Equation \eqref{eq:CME} compares to exact
results derived for two specific model well studied in the literature. We show that the mean-field (fully connected) ferromagnet
is well described with the help of the CME. We also use the exactly solvable one dimensional lattice to illustrate
that \eqref{eq:CME} gives the right result in a low correlation limit.

\subsection{Mean field ferromagnet with Glauber dynamics}

We start considering the Cavity Master Equation equation \eqref{eq:CME} and assuming that the Glauber transition rate \cite{Glauber63}
\begin{equation}
r_i(\sigma_i,\sigma_{\partial i}) = \frac{\alpha}{2} (1 - \sigma_i \tanh( \beta \frac{J}{N} \sum_{k \neq i} \sigma_k)
\label{eq:glauber}
\end{equation}
where all the spins are consider to be at time $t$. One can then introduce a delta function for the variable $m = \frac{1}{N} \sum_{k \neq i} \sigma_k$, and use its integral representation to rewrite the RHS of (\ref{eq:CME}) as:
\begin{align}
 \nonumber
 \dot{p}(\s_i,t|\s_j) =& -
\int d\hat{m} \, dm\,  e^{-i m \hat{m}} \, r_i(\sigma_i,m) \, e^{i \frac{ m \hat{m}}{N} \sigma_j(t)} \big[ \sum_{\sigma_k} e^{i \frac{ m \hat{m}}{N} \sigma_k(t)} p(\sigma_k| \sigma_i) \big]^N p(\sigma_i,\sigma_j)\\ &
+ \int d\hat{m}\, dm\,  e^{-i m \hat{m}} r_i(\sigma_i,m) e^{i \frac{ m \hat{m}}{N} \sigma_j(t)} \big[ \sum_{\sigma_k} e^{i \frac{ m \hat{m}}{N} \sigma_k(t)} p(\sigma_k| -\sigma_i) \big]^N p(-\sigma_i,\sigma_j)
\label{eqn:meanfieldCMEtwo}
\end{align}
Note now that the term within brackets can be expanded as:

\begin{equation}
\big[ \sum_{\sigma_k} e^{i \frac{ m \hat{m}}{N} \sigma_k(t)} p(\sigma_k| \sigma_i) \big]^N = \sum_k C_{N,k} e^{i \hat{m} (1 - \frac{2k }{N}) } p^k(1|\sigma_i)  p^{N-k}(-1|\sigma_i)
\label{eq:brack}
\end{equation}
\noindent that substituted in (\ref{eqn:meanfieldCMEtwo}) allows a simple integrationg over $\hat{m}$ such that now:
\begin{align}
 \nonumber
 \dot{p}(\s_i,t|\s_j) =& -
\int dm(t) P(m(t) | \sigma_i, \sigma_j)  r_i(\sigma_i,m)  p(\sigma_i,\sigma_j)\\
&+ \int dm(t) P(m(t) | -\sigma_i, \sigma_j)  r_i(-\sigma_i,m)  p(-\sigma_i,\sigma_j)
\label{eqn:meanfieldCMEthree}
\end{align}
\noindent where $P(m(t) | \sigma_i, \sigma_j) = \sum_k C_{N,k} 
p^k(1|\sigma_i)  p^{N-k}(-1|\sigma_i) \delta( m(t) - (1- \frac{2 k}{N}) -\frac{\sigma_j(t)}{N})$.

One can then define $m_i = \sum_{\sigma_i} p(\sigma_i|\sigma_j) \sigma_i$ as the magnetization of spin $i$ at time $t$, provided that the spin $j$ is in state $\sigma_j$. It is then straighforward to show that:
\begin{equation}
\dot{m}_i = - \alpha \, m_i(t) + \alpha\int dm(t) \,P(m(t)| \sigma_j) \, \tanh(\beta J m(t))
\label{eq:semiLang}
\end{equation}
A further average of (\ref{eq:semiLang}) over the single spins then gives:

\begin{equation}
\dot{m}(t) = - \alpha \,m(t) + \alpha \int dm(t) \, P(m(t)| \sigma_j) \, \tanh(\beta J m(t))
\label{eq:}
\end{equation}

\noindent that looks similar to the standard results of the literature, altough it may have a more subtle interpretation due to the conditional distribution on the spin $\sigma_j$. One can proceed assuming that in the thermodynamics limit the importance of the state of one individual spin at time $t$ is neglegible and that the average magnetization is defined by only one possible trajectory, recovering the more standard result:
\begin{equation}
\dot{m}(t) = - \alpha\,  m(t) + \alpha \, \tanh(\beta J m(t))
\label{eq:}
\end{equation}

\subsection{One dimensional ferromagnet with Glauber dynamics}

One dimensional lattices are often a good starting point to seach for analytical results that could be used to understand the nature and limits of approximations made in more involved contexts. In this subsection we consider the 1D ferromagnet using the CME \eqref{eq:CME} and compare this
solution to Glauber's exact result \cite{Glauber63}.

Let us start by quoting the exact results. Observe that the transition rate \eqref{eq:glauber} can be written for this 1D case as
\begin{equation}
r_i(\sigma_i,\sigma_{\partial i}) = \frac{\alpha}{2} \big(1 - \sigma_i \tanh( \beta J (\s_{i-1}+\s_{i+1})\big) =  \frac{\alpha}{2} \big(1 - \frac{\sigma_i}{2} (\s_{i-1}+\s_{i+1}) \tanh(2 \beta J) \big)
\label{eq:glauber1D}
\end{equation}
Note that we only have two neighbours of spin $i$, therefore $\p i = \{i-1, i+1\}$. This rate, put into the exact ME \eqref{eq:localME}, gives 
the following closed set of equations for the single site magnetizations:

\begin{align}
\dot{m_i} = - \alpha \, m_i + \frac{\alpha}{2} \tanh(2 \beta J) \Big(m_{i-1} + m_{i+1}\Big).
\label{eq:miME}
\end{align}
and equivalently for the probabilities:
\begin{eqnarray}
 \dot p(\s_{i}) &=& A(m_{i},m_{i-1},m_{i+1}) \s_{i}\\
  &=& - \dfrac{\alpha }{2} \left[ m_{i} - \dfrac{\tanh (2\beta J)}{2} (m_{i-1} + m_{i+1})\right]\s_{i}
  \label{eq:MEexact1D}
\end{eqnarray}

If we now average \eqref{eq:miME} over all the sites $i$ in the graph, in order to get the equation for the global magnetization, we obtain
\begin{align}
\dot{m} = - \alpha \,\Big( 1 - \tanh(2 \beta J)\Big) m
\label{eq:exactGlaubglobal}
\end{align}
whose solution is $ m(t) = m(0) e^{-\gamma \,t }$, with $\gamma = (1 - \tanh(2 \beta J))$. Equations \eqref{eq:miME}, \eqref{eq:MEexact1D} and 
\eqref{eq:exactGlaubglobal} correspond to the exact results found by Glauber for the evolution of the magnetization of an Ising chain.

Now let us see what would be the dynamics if instead of the exact ME we considered the Cavity Master Equation equation \eqref{eq:CME} as a starting point in our analysis. Substituting in \eqref{eq:CME} the transition rate expressed in \eqref{eq:glauber1D} reads
\begin{align}
 \nonumber
 \frac{d p(\s_i|\s_{i+1})}{d t} = -  \sum_{\sigma_{i-1}} \Bigg[& \frac{\alpha}{2} \big(1 - \frac{\sigma_i}{2} (\s_{i-1}+\s_{i+1}) \tanh(2 \beta J) \big) p(\sigma_{i-1} |\s_i) p(\s_i|\s_{i+1}) \\
 &-  \frac{\alpha}{2} \big(1 + \frac{\sigma_i}{2} (\s_{i-1}+\s_{i+1}) \tanh(2 \beta J) \big) p(\sigma_{i-1} |-\s_i) p(-\s_i |\s_{i+1}) \Bigg]
\label{eq:CME_1D}
\end{align}
Now we can multiply both side of the above equation by $\s_i$ and marginalize over it. Similarly to what is done  for the mean-field ferromagnet, we define $m_i(\s_{i+1}) = \sum_{\sigma_i} p(\sigma_i|\sigma_{i+1}) \sigma_i$ to be the magnetization of spin $i$ at time $t$, given that the spin $i+1$ is in state $\sigma_{i+1}$. Separating terms the above equation then reads
\begin{align}
 \nonumber
\dot{m_i}(\s_{i+1}) =& -  \sum_{\s_i, \,\sigma_{i-1}} \Bigg[ \frac{\alpha}{2}  \,\s_i \Big(p(\sigma_{i-1} |\s_i) p(\s_i|\s_{i+1}) -p(\sigma_{i-1} |- \s_i) p(- \s_i|\s_{i+1})\Big)\\
&- \frac{\alpha}{4} \tanh(2 \beta J)(\s_{i-1}+\s_{i+1})\Big( p(\sigma_{i-1} |\s_i) p(\s_i|\s_{i+1}) + p(\sigma_{i-1} |-\s_i) p(-\s_i |\s_{i+1})\Big) \Bigg]
\label{eq:CME_1D_sep}
\end{align}

Marginalizing we get
\begin{align}
 \nonumber
\dot{m_i}(\s_{i+1}) =& - \alpha \, m_i(\s_{i+1}) \\ 
&+ \frac{\alpha}{2} \tanh(2 \beta J) \big( m_{i-1}(\s_{i+1}) + \s_{i+1} \big)
\label{eq:CME_1D_sepb}
\end{align}

This approximate equation for the local magnetization reflects the structure of the equation (\ref{eq:miME}) deduced starting from the Master Equation, but has a different meaning. As already said the quantity $ m_i(\s_{i+1})$ should be interpreted as the magnetization of spin $i$, {\em conditioned} to spin $i+1$ being in state $\s_{i+1}$. 

To compare to the exact results we need the real magnetization $m_i$. It is simply related to $m_i(\s_{i+1})$ through $m_i = \sum_{\s_{i+1}} m_i(\s_{i+1}) p(\s_{i+1})$. The time derivative of $m_i$ is:

\begin{equation}
 \dot m_i = \sum_{\s_{i+1}} \left[ \dot m_i(\s_{i+1}) p(\s_{i+1}) +  m_i(\s_{i+1}) \dot p(\s_{i+1}) \right]
 \label{eqn:fullmag}
\end{equation}

We can use \eqref{eq:CME_1D_sepb} and \eqref{eq:MEexact1D} in the equation above and obtain:

\begin{eqnarray}
 \nonumber
 \dot m_i &=& -\alpha m_i + \dfrac{\alpha}{2} \tanh(2\beta J) (m_{i-1} + m_{i+1})\\
 &+& A(m_{i},m_{i+1},m_{i+2}) [m_i(\s_{i+1}=1) - m_i(\s_{i+1}=-1)]
 \label{eqn:finalresult}
\end{eqnarray} 

The first line in (\ref{eqn:finalresult}) is the exact result. The second line is an extra term that goes to zero for weak correlation or high temperature because $m_i(\s_{i+1})$ becomes independent of $\s_{i+1}$. Therefore we can
 conclude that our CME is consistent with exact results when appropiate limits are taken. For models with higher connectivities
 we may expect some agreement with numerical simulations even for not so high temperatures since a higher number of neighbours
 translates directly into a weaker correlation between spins.

\section{Numerical results}\label{sec:num_results}

To test numerically the performance of the approximate dynamics described by \eqref{eq:localMEfact} and (\ref{eqn:alltogether3_approx})we compare the numerical solutions of this set of equations with the results obtained after running a large number of continuous
time kinetic Monte Carlo simulations. 

Three typical models are considered: an Ising ferromagnet with zero external magnetic field, the same ferromagnet with disordered local magnetic field (also known as Random Field Ising Model, RFIM) and the Viana-Bray spin-glass model, where interaction constants $J_{ij}$ are drawn positive or negative with equal probability from a bimodal distribution. All three systems  share the same underlying topology of an instance of an Erd\"os-R\'enyi random graph, generated with $N=1000$ nodes and mean connectivity $c=3$. 
The rate of change for individual spins is taken according to Glauber's rule:
\begin{equation}
 r_{i}(\s_i,\s_{\partial i}) = \dfrac{\alpha}{2} (1-\s_i \tanh[\beta(\sum_{j\in 
\partial i}J_{ij}\s_j + h_i)])
\end{equation}
\noindent where $\alpha$ is a constant defining the time unit, $t_0=1/\alpha$. For the actual simulations, the interaction
constants are rescaled by a factor $1/c$.
%
%
%
\begin{figure}[t!]  
\begin{center}
        \subfloat[Ising ferromagnet]{\includegraphics[height= 3.9cm]{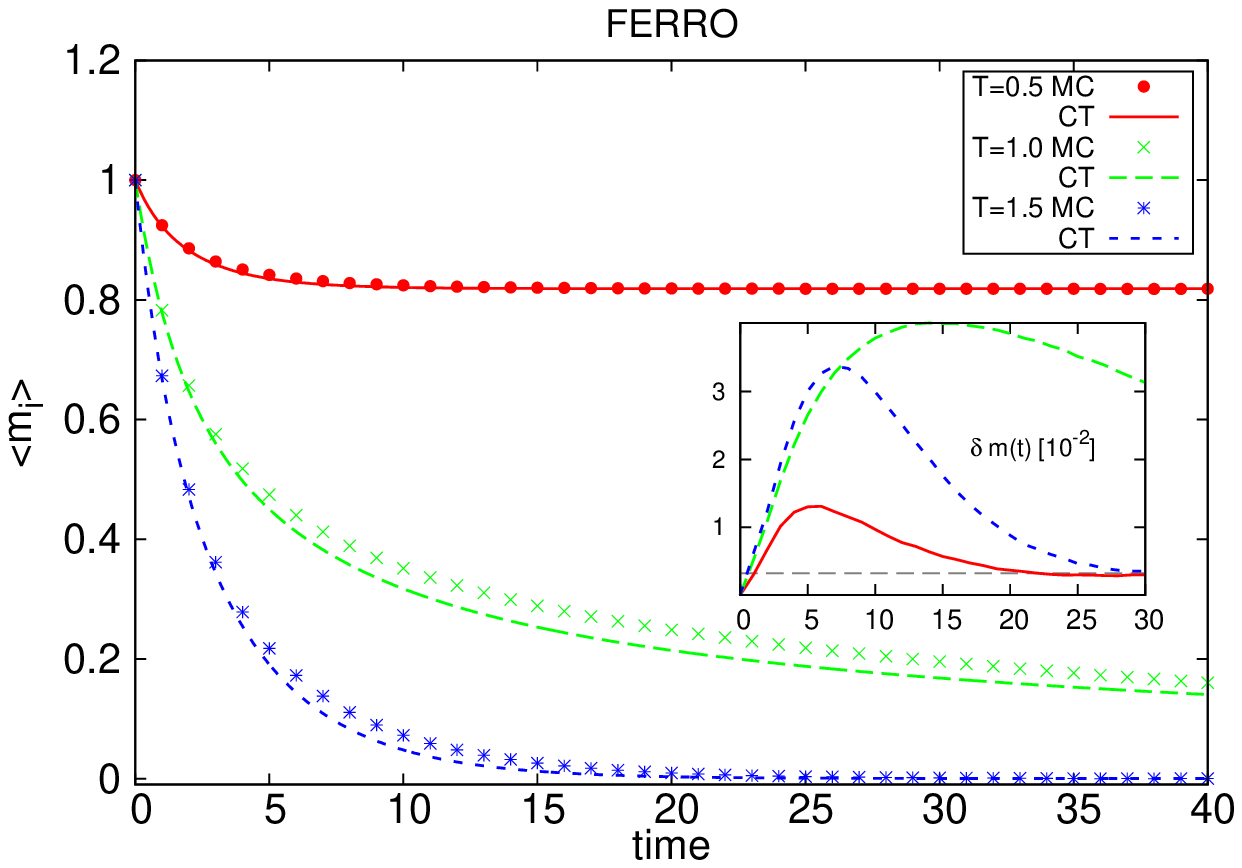}\label{fig:magn_ferro}}
\subfloat[RFIM]{\includegraphics[height= 3.9cm]{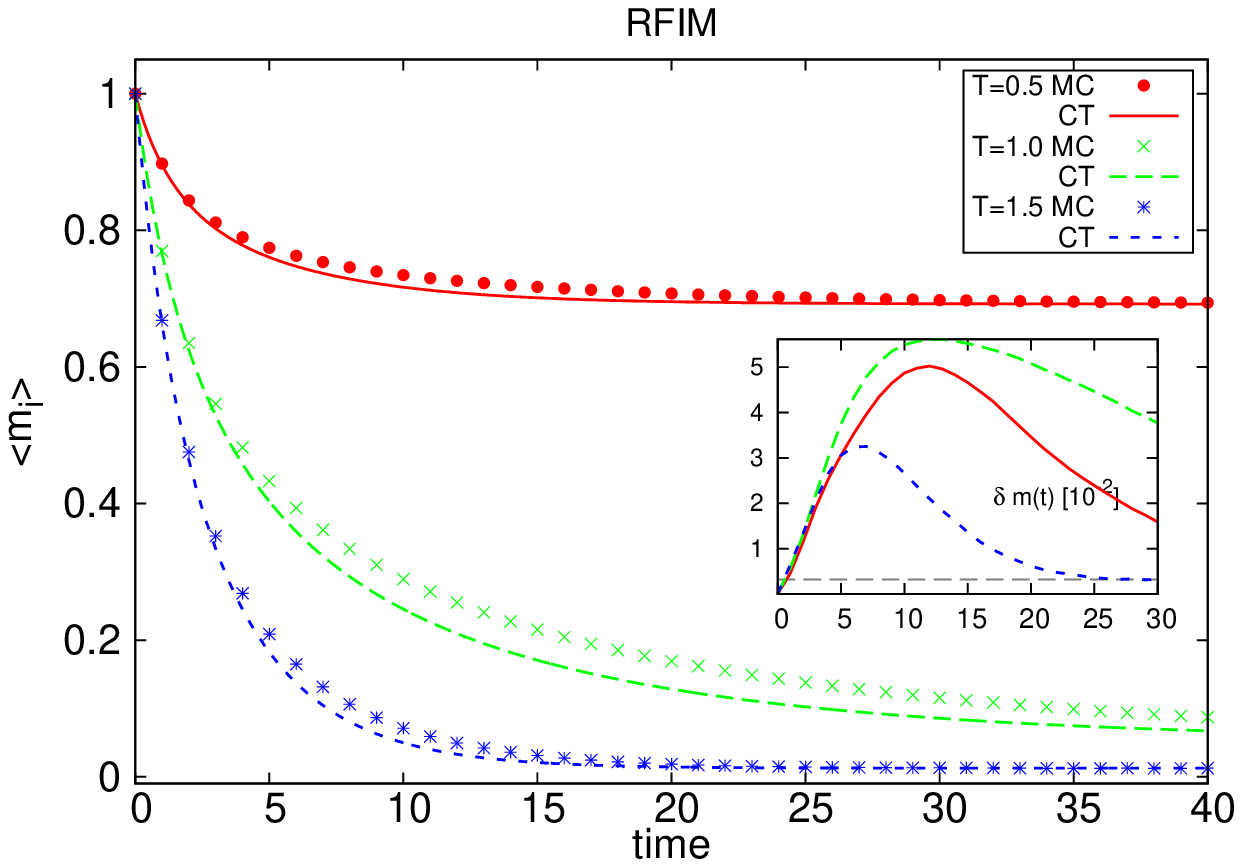}\label{fig:magn_RFIM}}
\subfloat[Spin Glass]{\includegraphics[height= 3.9cm]{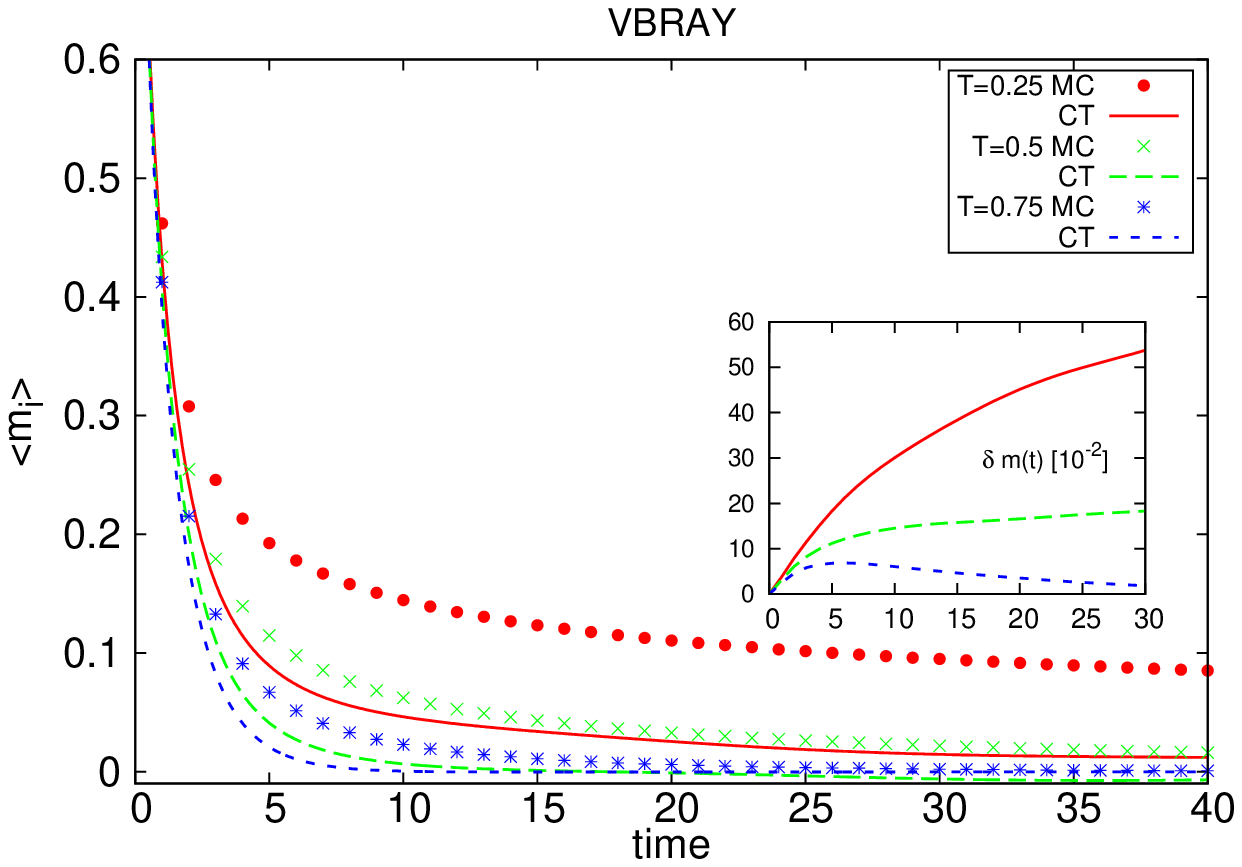}\label{fig:magn_SG}}
 \caption{\small Evolutionary dynamics of the global magnetization parameter for three models of spins in a random graph: the Ising ferromagnet, the random field Ising model and the Viana-Bray Model. The insets show the mean error in local magnetization, as defined in the main text. Dynamic cavity works for ferromagnets and RFIM but fails for the spin glass model.}
 \label{fig:global_mag_sym}
   \end{center}
\end{figure}

The CME for the conditional probabilities (\ref{eqn:alltogether3_approx})
is solved using Euler's method for ODEs. The integration stepsize $h$ is
 a fraction of this time unit, $\Delta t=0.05\, t_0$. Initial conditions, both for the differential equations and the stochastic simulations, correspond to a frozen state with all spins pointing in the same (positive) direction. At finite temperature this is not an equilibrium state and the system will evolve
and relax towards it.
Once integrated, the deduced equations give the time evolution of conditional pairwise probabilities, but for observables
we need complete probability distributions. These can then be obtained
integrating the factorized master equation for a local variable  (\ref{eq:localMEfact}) . where, the conditional probabilities that appear in the previous equation are given by the solution of (\ref{eqn:alltogether3_approx}). 

Starting with local probability distributions, local magnetizations are defined as usual, $m_i(t)=\sum_{\s_i(t)} \s_i(t) \, p(\s_i(t))$, where $p(\s_i(t))$ is estimated by (\ref{eq:localMEfact}) and (\ref{eqn:alltogether3_approx}). Global
magnetization is computed as the average of local ones over the network $m(t)=\frac{1}{N}\sum_i m_i(t)$. For disordered systems it is also useful to investigate the evolution of the Edwards-Anderson (EA) parameter, defined as the average of the squared local magnetization $q_{EA}(t)=\frac{1}{N}\sum_i m_i^2(t)$.

%
%
\begin{figure}[t!]
%
        \subfloat[Ising Ferromagnet]{\includegraphics[height= 3.9cm]{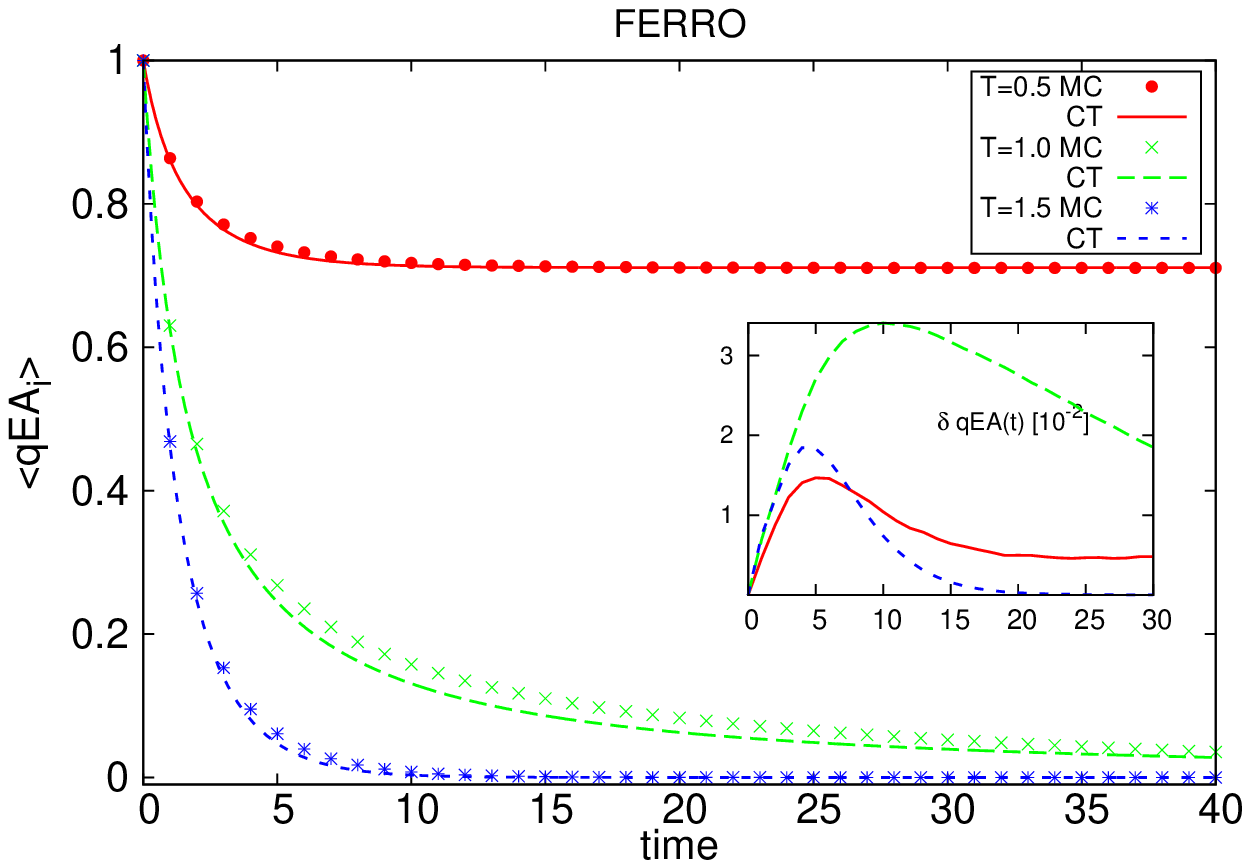}\label{fig:q_ferro}}
\subfloat[RFIM]{\includegraphics[height= 3.9cm]{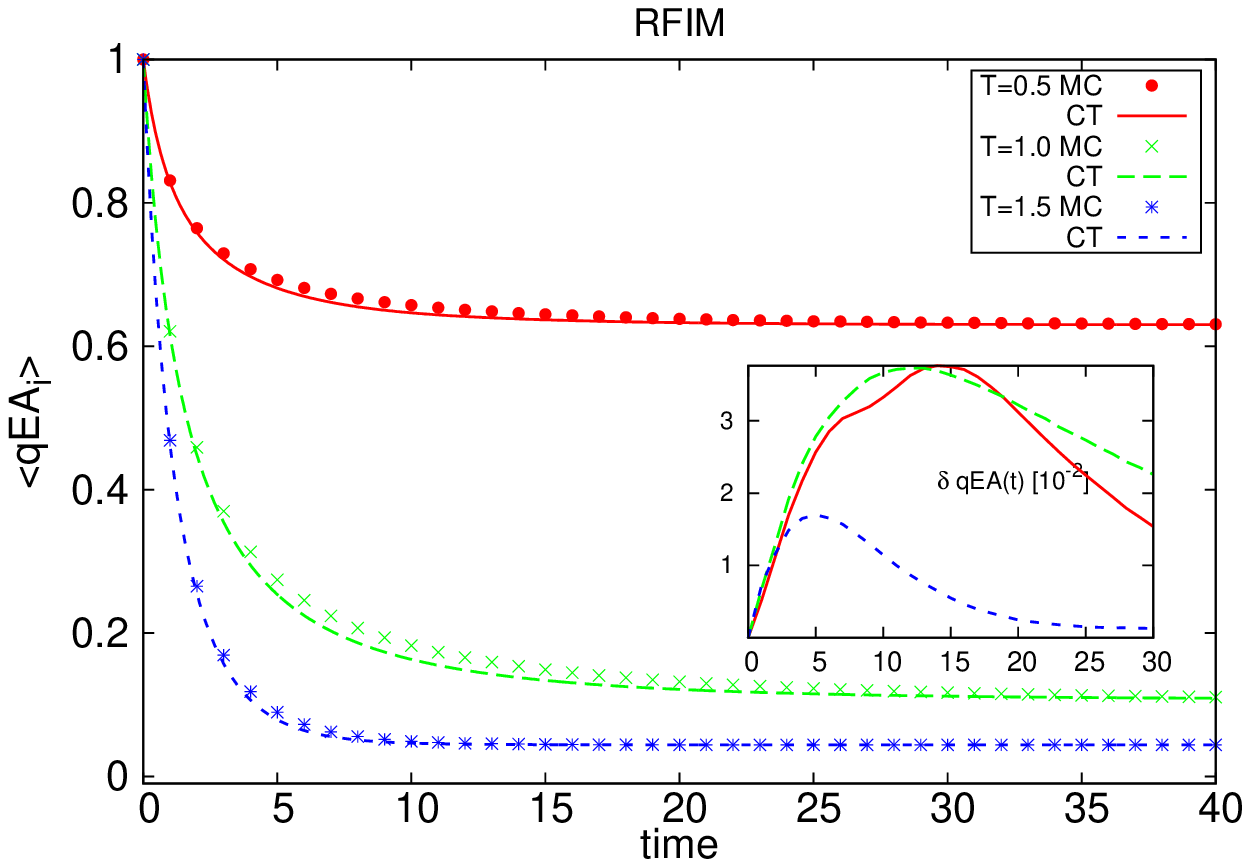}\label{fig:q_RFIM}}
\subfloat[Spin Glass]{\includegraphics[height= 3.9cm]{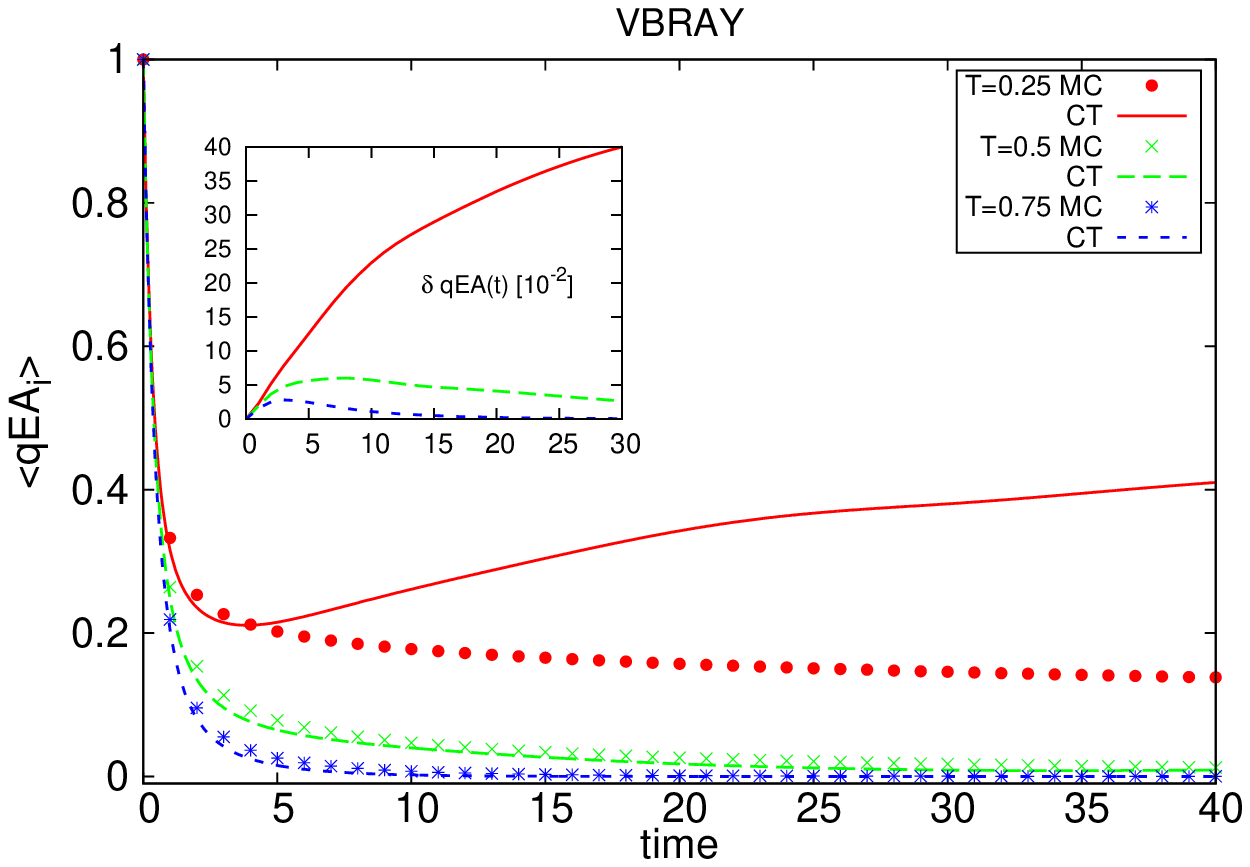}\label{fig:q_SG}}
        \caption{\label{fig:global_qEA_sym} \small Dynamics of the global EA parameter. The behavior for low temperatures in the Viana-Bray model shows that
        even though global magnetization is close to zero for long times (see Fig. \ref{fig:magn_SG}), spins are locally magnetized for long times, as it is expressed by the non zero value of $qEA(t)$ for $T=0.25$.}
\end{figure}

Figures \ref{fig:global_mag_sym} and \ref{fig:global_qEA_sym} show the relaxation of global magnetization $m(t)$ and $q_{EA}(t)$ for our three test models, using MC simulations (dots)
and the CME formalism (lines). The insets include the mean error of local magnetization with respect to the MC predictions, $\delta m=\sqrt{\frac{1}{N}\sum_{i}(m^{CME}_i(t)-m^{MC}_i(t))^2}$. For the EA parameter we define the equivalent error measure. 

The ferromagnetic case shown in Figure \ref{fig:magn_ferro} and \ref{fig:q_ferro} displays a good agreement for the transient regime as well as for the long time behaviour for
temperatures above and below the critical $T_c\approx0.96$ for this model. For a value $T=1$, quite close to the phase transition,
the qualitative behaviour of the magnetization dynamics is fairly reproduced but its accuracy, as expected, diminishes. An important part of the procedure leading to
(\ref{eqn:alltogether3_approx}) relies on the factorization of distributions and this is equivalent to set almost all (connected) correlations to zero. It is therefore natural to find a failure nearby a region where correlations become fundamental, as it is the case for a second order phase transition.

For the RFIM, which is one of the standard literature examples of disordered system, the dynamic cavity equation reproduces the dynamical behaviour with a quality comparable to the ferromagnetic case, see Figure \ref{fig:magn_RFIM} and \ref{fig:q_RFIM}. Errors found in this case are of the same order of magnitude to the situation where $h_i=0$.

The Viana-Bray model, on the other hand, shows errors one order of magnitude larger than the previous models, which worsen as temperature decreases, see Figure \ref{fig:magn_SG} and \ref{fig:q_SG}.
It is known that this model has a spin glass 1RSB transition and this implies a fundamental difference with respect to the previously considered models. The state space collapses
for low temperature into a hierarchy of low energy configurations and it is no longer well described by only one equilibrium solution.

\begin{figure}[t!]
        \subfloat[Ising Ferromagnet]{\includegraphics[height= 3.9cm]{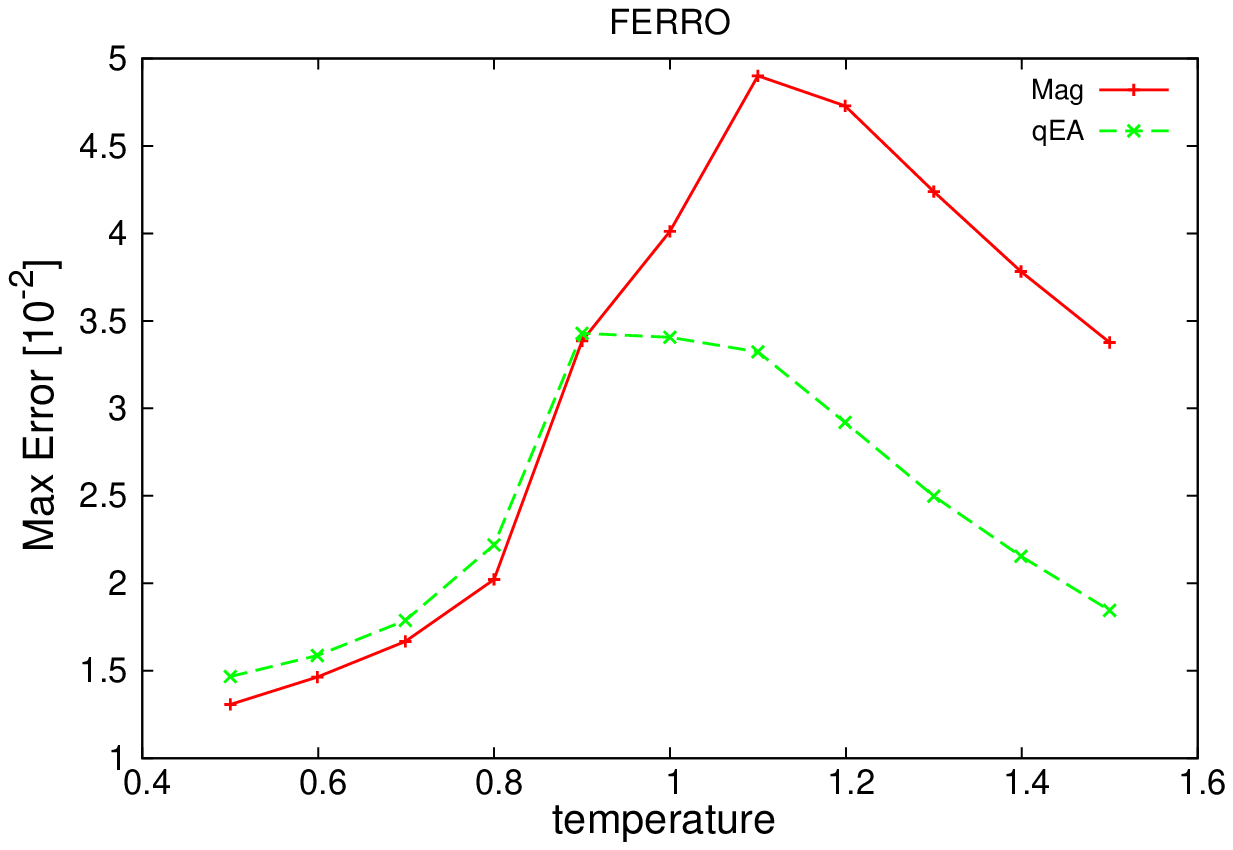}\label{fig:err_ferro}}
\subfloat[RFIM]{\includegraphics[height= 3.9cm]{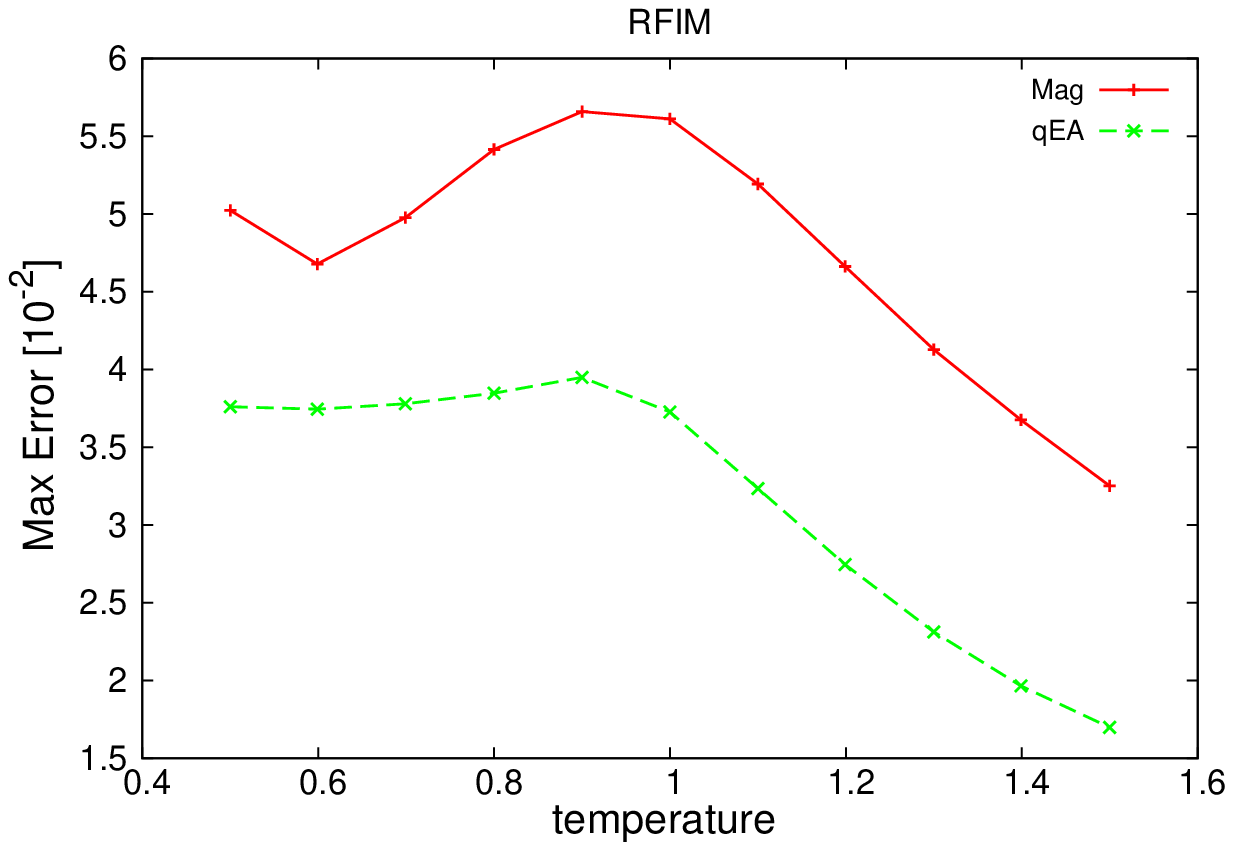}\label{fig:err_RFIM}}
\subfloat[Spin Glass]{\includegraphics[height= 3.9cm]{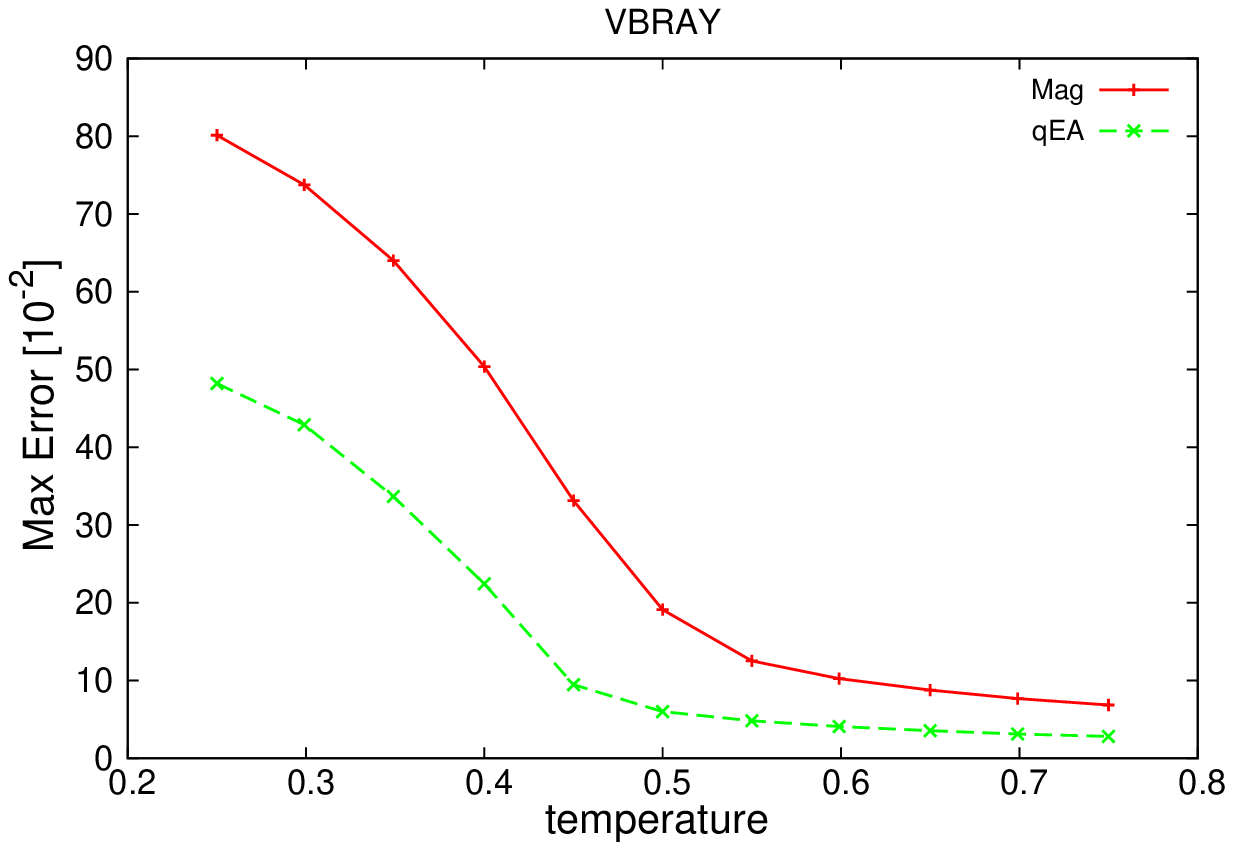}\label{fig:err_SG}}
        \caption{\label{fig:max_error_vs_temp} \small Maximum mean error dependence with temperature. For the ferromagnet and the RFIM, error
        increases before and decreases after the ferromagnetic transition as the temperature changes. For the Viana-Bray spin glass, error
        grows monotonically lowering the temperature.}
\end{figure}

In Figure \ref{fig:max_error_vs_temp} we present, for all models,  the temperature dependence of the maximum value of the mean error. Note that, in the insets of Figure
\ref{fig:global_mag_sym} and \ref{fig:global_qEA_sym}, the error is maximum 
for an intermediate time in the ferromagnet and the RFIM. It is important to notice that in all cases the errors are low at very short times as well as in the stationary regime. In the SG model, though, the error increases
monotonically with time for low temperatures in opposition to what happens with the ferromagnet and RFIM. As we said before, this indicates that there is a wrong assumption regarding the structure of phase space.
%
%

\section{Conclusions}

In this work we have derived a new method to close the Master Equation for the continuous dynamics of interacting spins. The approach relies on the factorization of the conditional distribution of the state of spin $i$ and its neighbours and on the formalism of the theory of Random Point Processes. By assuming a tree-like graph topology,  using this formalism, we are able to re-derive a known equation for conditional probabilities of spin histories which is called dynamic message-passing or dynamic cavity equation in the literature.  Combining this result with the approximated master equation and using the Random 
Point Process formalism, we are able to parametrize probability distributions of the spin histories and obtain a rigorous derivation of a new dynamic equation for the conditional distribution of spin variables, the Cavity Master Equation. This new equation together with the Master Equation for the single spin dynamics completely determine the temporal evolution of the model. 

We have shown that our approach reproduces the known analytical solution of two prototypical models and we have tested numerically the performances of the method for three more complex cases defined on Random Graphs. Numerical results show a quantitative good agreement with Monte Carlo simulations for those models which do not have a spin glass phase. For the Viana-Bray model, the technique fails below the glass transition.

We believe that the general nature of our method allows to apply it on models with different transition rates, networks with various connectivity symmetries (asymmetric and partially symmetric graphs) and therefore could bring to several further developments to investigate the dynamics of physical and biological systems. Extension to models with a glassy phase is under development.

\textbf{Acknowledgements.} This research is supported by the Swedish Science Council through grant 621-2012-2982 and by the Academy of Finland through its Center of Excellence COIN (EA), by European Union through Marie Curie ITN ``NETADIS'' FP7/2007-2013/grant agreement n. 290038 (GDF), by STINT and EUSBIOS.

\bibliographystyle{unsrt}
\bibliography{general}

\end{document}